\documentstyle[12pt]{article}

\newcommand{\be}{\begin{equation}}
\newcommand{\ee}{\end{equation}}
\newcommand{\dd}{\partial}
\newcommand{\bea}{\begin{eqnarray}}
\newcommand{\eea}{\end{eqnarray}}

\begin{document}
\baselineskip .25in
\newcommand{\numero}{hep-th/9709071}  

\newcommand{\titre}{Non-Abelian Duality Based on Non-Semi-Simple Isometry 
Groups}
\newcommand{\auteura}{ } 
\newcommand{\auteurb}{Noureddine Mohammedi}

\newcommand{\placea}{Laboratoire de Math\'ematique et Physique Th\'eorique
CNRS/UPRES-A 6083,\\Universit\'e Fran{\c{c}}ois Rabelais\\
Facult\'e des Sciences et Techniques\\
Parc de Grandmont\\
F-37000 Tours, France}
\newcommand{\placeb}{Laboratoire de Math. et Physique Th\'{e}orique,
Universit\'{e} de Tours, Parc de Grandmont,F-37200
 Tours,France.}
\newcommand{\beq}{\begin{equation}}
\newcommand{\eeq}{\end{equation}}

\newcommand{\abstrait}
{Non-Abelian duality transformations built on non-semi-simple isometry 
groups are analysed. We first give the conditions under which the original
non-linear sigma model and its non-Abelian dual are equivalent. The existence
of an invariant and non-degenerate bilinear form for the isometry Lie algebra 
is crucial for this equivalence. The non-Abelian dual of a conformally invariant
sigma model, with non-semi-simple isometries, is then constructed and its beta 
functions are shown to vanish. This  study resolves an apparent obstruction 
to the conformal invariance of sigma models obtained via non-Abelian duality 
based 
on non-semi-simple groups.}
\begin{titlepage}
\hfill \numero  \\
\vspace{.5in}
\begin{center}
{\large{\bf \titre }}
\bigskip \\ \auteura 
\bigskip  {}
\auteurb  \footnote{e-mail:
nouri@celfi.phys.univ-tours.fr}
\bigskip \\ 
\placea  \bigskip \\
\vspace{.9 in} 
{\bf Abstract}
\end{center}
\abstrait
 \bigskip \\
\end{titlepage}
\newpage

\section{Introduction}

Duality transformations have been the centre of wide 
investigations recently.
(see \cite{amit,loza1} for a review and further references).
They are crucial tools for understanding
the structure of the moduli space resulting from 
compactifications of string theories. 
These transformations connect two apparently different
conformal string backgrounds. It is in this sense 
that they are going to be treated in here. We will not be concerned
with the global issues or the reversibility of these transformations.
The implementation of duality relies heavily on the existence of isometries 
on target space-time backgrounds \cite{rocek}.
However, duality can also be defined 
without isometries. The latter case is known
as Poisson-Lie T-duality \cite{klimcik}.
\par
The duality is Abelian if it is built on an Abelian isometry
group \cite{buscher,abelian} and non-Abelian when its corresponding 
isometry group is non-Abelian \cite{nonabelian}. 
In both cases, the starting point is a 
non-linear sigma model enjoying some global symmetries which form 
some isometry groups \cite{rocek}.
The dual sigma model is then obtained by gauging one (or more) 
isometry groups
and at the 
same time constraining, by means of Lagrange multipliers, the field 
strength of each gauge field to vanish. Integrating out these 
Lagrange multipliers and fixing the gauge invariance yields the original
model. On the other hand, integrating over the gauge fields and keeping
the Lagrange multipliers results in the dual sigma model.  
One expects then that
if the original backgrounds are conformal backgrounds (i.e. they satisfy
the vanishing of the beta functions conditions) then the dual backgrounds
are also conformal and form consistent backgrounds on which the string 
propagates \cite{buscher,rocek,haag}.
\par
This is indeed the case when dealing with Abelian duality. However,
for non-Abelian duality some obstruction to the conformal invariance
of the dual theory have been reported. The authors of \cite{venezia}
considered
a cosmological solution to string theory (the beta functions vanish)
based on a Bianchi V space-time. The application of non-Abelian 
duality transformations to this string solution does not lead
to conformal string backgrounds. A similar conclusion was reached 
when considering a string solution based on
Bianchi IV \cite{lozano}. 
Both examples, however,  share the same feature: their 
non-Abelian isometry groups are both non-semi-simple.
\par
We examine, in this letter, the implementation of non-Abelian
duality when the gauged isometry group is non-semi-simple.
We show that the Lagrange multiplier, necessary for forcing the
gauge field to be a pure gauge, fulfills his r\^ole only 
when the Lie algebra of the isometry group is endowed with
an invariant and non-degenerate bilinear form. In other
words, the original sigma model and its gauged version 
(with the Lagrange multiplier term) are equivalent only if
the invariant bilinear form is invertible. The above mentioned
examples, where non-Abelian duality seems to fail,  do not possess
an invertible bilinear form. 
\par
We provide here a two-dimensional non-linear sigma model
having a non-semi-simple isometry group whose Lie algebra
possesses an invariant bilinear form. The model is a 
WZW model based on the centrally extended Euclidean group $E^c_2$.
It is therefore conformally invariant to all order in perturbation theory.
We construct then the non-Abelian dual theory and explicitly check that its 
one-loop beta functions vanish. This model is presented in section 3.
The general procedure for obtaining dual sigma models is briefly
outlined in section 2. The crucial function of the invariant bilinear
form is also emphasized.  

\section{The duality procedure}

The original theory is given by the general ungauged bosonic two-dimensional 
non-linear sigma model
\bea
S\left(\varphi\right)&=&\int_{\dd \Sigma} {\mathrm {d}}^2x \sqrt
{\gamma}\left(
\gamma^{\mu\nu}G_{ij}\left(\varphi\right)\dd_\mu\varphi^i\dd_\nu\varphi^j
+ \Phi\left(\varphi\right)R^{(2)}\right)
+\Gamma\left(\varphi\right)\nonumber\\
\Gamma\left(\varphi\right)&=&
\int_{\Sigma}{\mathrm {d}}^3y{\epsilon}^{\mu\nu\rho}H_{ijk}\left(\varphi\right)
\dd_\mu\varphi^i\dd_\nu\varphi^j\dd_\rho\varphi^k\,\,\,.
\label{ungaugedaction}
\eea
In this equation $\gamma_{\mu\nu}$ is the metric on the two-dimensional
world sheet $\dd \Sigma$, $\gamma$ is its determinant and $R^{(2)}$ is the 
scalar
curvature. The metric $G_{ij}$ and the dilaton field $\Phi$ correspond to 
the set of massless modes of an associated string theory. The other massless
field, the anti-symmetric tensor $B_{ij}$, is defined through its three-form
$2H_{ijk}=\dd_i B_{jk}+\dd_k B_{ij} +\dd_j B_{ki}$. 
The latter is defined on the tree-dimensional space $\Sigma$ whose
boundary is $\dd \Sigma$.
\par
The sigma model Lagrangian is manifestly invariant under global 
reparametrisation of the target space-time. However, the only global symmetries 
suitable for gauging are those for which the metric remains form invariant. 
Such symmetries form a Lie group $\mathcal{G}$, namely the 
isometry group of the metric $G_{ij}$. 
A general infinitesimal isometry on the 
target space-time is given by the global transformation
\be
\delta \varphi^i=\alpha^a K^i_a\left(\varphi\right) \,\,\,,
\label{phi}
\ee
where $\alpha^a$ is a constant.
The generator of this transformation are 
$K_a=K^i_a\dd_i\,\,$ and they satisfy the
Lie algebra of $\mathcal{G}$
\be
\left[K_a\,\,,\,\,K_b\right]=f^c_{ab}K_c \,\,\,\,\,\,
{\mathrm{or}}\,\,\,\,\,\,
K^i_a\dd_iK_b^j-K^i_b\dd_iK^j_a=f^c_{ab}K^j_c\,\,\,,
\label{isometry}
\ee
where $f^a_{bc}$ are structure constants supposed to be field-independent.
\par
Global invariance of the action implies some constraints on 
$G_{ij}$, $H_{ijk}$ and $\Phi$. The condition on the metric term are simply
the Killing equations
\be
G_{ij}\nabla_k K^j_a + G_{kj}\nabla_i K^j_a=0\,\,\,,
\ee
where the covariant derivative $\nabla_i$ is with respect to the 
metric $G_{ij}$. The dilaton term is invariant when 
$K^i_a\dd_i \Phi =0$.
\par
The last term of the action is invariant under (\ref{phi}) provided that the
torsion $H_{ijk}$ obeys
\be
K^i_a\dd_i H_{jkl}  + H_{ikl}\dd_j K^i_a + H_{jil}\dd_k K^i_a 
+H_{jki}\dd_lK^i_a=0\,\,\,.
\label{Hcond}
\ee
Since the three-form $H_{ijk}$ is closed, the invariance condition 
requires, for every vector $K^i_a$, the existence of a globally defined 
one-form fulfilling
\be
K^i_a H_{ijk} =\dd_j L_{ak}-\dd_k L_{aj}\,\,\,.
\label{Ldef}
\ee
This last equation is at the heart of gauging the general sigma model.
\par
The first step towards constructing the dual theory is to gauge 
the above global symmetry. We therefore introduce a gauge field
$A_\mu^a$ taking values in the Lie algebra ${\cal {G}}$ and transforming
as 
\be
\delta A^a_\mu=-\dd_\mu\alpha^a - f^a_{bc}A^b_\mu\alpha^c_\mu\,\,\,\,.
\label{gaugevar}
\ee
The gauging is then possible only if we impose two further conditions
\cite{ian,chris}
\bea
K^i_a\dd_iL_{bj} + L_{bi}\dd_jK^i_a =-f^c_{ab}L_{cj}\nonumber\\
L_{ai}K^i_b + L_{bi}K^i_a=0 
\,\,\,.
\label{rep}
\eea
The last condition ensures that the gauge fields appear at most
in a quadratic form and live entirely on the two-dimensional manifold.
\par
The gauged action is then found to be given by
\bea
S_{\rm{gauge}}\left(\varphi,A\right)&=&
\int_{\dd \Sigma} {\mathrm {d}}^2x \sqrt{\gamma}\left(
\gamma^{\mu\nu}G_{ij}\left(\varphi\right)D_\mu\varphi^i D_\nu\varphi^j
+ \Phi\left(\varphi\right)R^{(2)}\right)
\nonumber\\
&-& 6 \int_{\dd \Sigma}{{\mathrm{d}}}^2x 
\epsilon^{\mu\nu}\left(
L_{ai} A^a_\mu\dd_\nu\varphi^i
+{1\over 4} 
\left(L_{ai}K^i_b-L_{bi}K^i_a\right)
A^a_\mu A^b_\nu\right)\nonumber\\
&+&\int_{\Sigma}{{\mathrm {d}}}^3y{\epsilon}^{\mu\nu\rho}
\left[H_{ijk}\left(\varphi\right)
\dd_\mu\varphi^i\dd_\nu\varphi^j\dd_\rho\varphi^k
\right] \,\,\,.
\eea
The covariant derivative is such that $D_\mu\varphi^i=
\dd_\mu\varphi^i +A_\mu^aK^i_a$.
\par
The dual theory is then constructed by considering the 
first order action given by \cite{rocek}
\be
S_1\left(\varphi,A,X\right)=S_{\rm{gauge}}\left(\varphi,A\right)
+ \int_{\dd \Sigma} {\mathrm {d}}^2x \epsilon^{\mu\nu}\Omega_{ab}
X^aF^b_{\mu\nu}\,\,\,.
\label{multiplier}
\ee
Here $F^a_{\mu\nu}=\dd_\mu A^a_\nu -\dd_\nu A_\mu
+f^a_{bc}A^b_\mu A^c_\nu$ is the field strenght and $X^a$ is the Lagrange 
multiplier. The Lagrange multiplier transforms as 
$\delta X^a=-f^a_{bc}X^b\alpha^c$
and takes values in the Lie algebra ${\cal G}$. That is $X=X^aT_a$
with $\left[T_a\,,\,T_b\right]=f^c_{ab}T_c\,$. 
The added term is gauge invariant provided that
\be
\Omega_{ac}f^c_{bd} +\Omega_{bc}f^c_{ad}=0 \,\,\,.
\label{form}
\ee
This means that $\Omega_{ab}$ is an invariant bilinear form of ${\cal G}$.
\par
The original theory is retrieved by integrating over the Lagrange
multiplier $X^a$. This integration leads, in the path
integral, to a delta function enforcing 
the condition 
\be
\Omega_{ab}F^b_{\mu\nu}=0\,\,\,.
\ee
This last equation yields $F^a_{\mu\nu}=0$ only if $\Omega_{ab}$ 
is invertible. In this case,
and only in this case,  that one can use gauge invariance to set $A^a_\mu$
to zero and hence to get the ungauged action.
One does not realise this issue by writing the Lagrange miltiplier term 
in the form $\epsilon^{\mu\nu}X_aF^a_{\mu\nu}$ \cite{venezia,lozano}. 
One must specify the 
invariant bilinear form used to lower the index of the Lagrange
multiplier $X^a$.
\par
The first example where non-Abelian duality failed was considered
by the authors of \cite{venezia}. 
There, a cosmological solution to string theory
in the form of a four-dimensional Bianchi V was analysed. The string
backgrounds contains a vanishing antisymmetric tensor field,
a constant dilaton and a metric in the form
\be
ds^2=-dt^2 +a^2\left(t\right)\left[dx^2+e^{-2x}
\left(dy^2+dz^2\right)\right]\,\,\,.
\ee
Conformal invariance then demands $a(t)=t$ and the metric becomes flat.
This metric possesses a non-Abelian isometry group generated by
\be
\left[K_1\,,\,K_2\right]=-K_{2} \,\,\,,\,\,\,
\left[K_1\,,\,K_3\right]=-K_{3} \,\,\,,\,\,\,
\left[K_2\,,\,K_3\right]=0\,\,\,.
\ee
which is realised by the differential operators
\be
K_1={\dd\over\dd x}+y{\dd\over\dd y}+z{\dd\over\dd z}\,\,\,,\,\,\,
K_2={\dd\over\dd y}\,\,\,,\,\,\,
K_3={\dd\over\dd z}\,\,\,.
\ee
The non-vanishing structure constants are $f^2_{12}=f^3_{13}=-1$.
The corresponding invariant bilinear form is found to be
\be
\Omega_{ab}=\left(
\begin{array}{ccc}
k&0&0\\
0&0&0\\
0&0&0\end{array}\right)\,\,\,,
\ee
where $k$ is an arbitrary constant. It is clear that this 
is degenerate. The integration over the Lagrange multiplier 
term in (\ref{multiplier}) leads to $F^1_{\mu\nu}=0$ only. 
This alone does not allow one
to set $A_\mu^a$ to zero and hence to arrive at the original theory.
Therefore, the first order action in (\ref{multiplier}) is not
equivalent to the original ungauged action. As a consequence, the 
action obtained by integrating out the gauge fields does not
necessarily provide conformal string backgrounds.
\par
A similar model was also studied in \cite{lozano}. 
It is based on the Bianchi IV type 
cosmological metric
\be
ds^2=-dt^2 +a^2\left(t\right)dx^2+b^2\left(t\right)e^{-x}
\left[dy^2+dz^2\right]\,\,\,.
\ee
With a zero torsion and a constant dilaton field, conformal invariance
at the one loop level imposes $a\left(t\right)=t/2$ and 
$b\left(t\right)=t$. This metric has also a non-Abelian isometry group
having the Lie algebra
\be
\left[K_1\,,\,K_2\right]=0 \,\,\,,\,\,\,
\left[K_2\,,\,K_3\right]=0 \,\,\,,\,\,\,
\left[K_1\,,\,K_3\right]=K_1\,\,\,
\ee
with the differential representation  
\be
K_1={\dd\over\dd y}\,\,\,,\,\,\,
K_2={\dd\over\dd z}\,\,\,,\,\,\,
K_3={\dd\over\dd x}+y{\dd\over\dd y}
\,\,\,.
\ee  
The only non-zero structure constants are $f^1_{13}=1$. Their 
unique invariant bilinear form is given by
\be
\Omega_{ab}=\left(
\begin{array}{ccc}
0&0&0\\
0&k&m\\
0&m&n\end{array}\right)\,\,\,
\ee  
with $k,m,n$ being arbitrary constants. Again this is not invertible
and the integration over the Lagrange multiplier 
gives $F^2_{\mu\nu}=F^3_{\mu\nu}=0$ when $\left(kn-m^2\right)\neq 0$.
This is not sufficient for obtaining the original theory.
\par
Therefore, the equivalence of the first order action  and 
the original sigma model can be established only when the isometry
group owns an invertible bilinear form. We present below a 
non-semi-simple isometry group having a non-degenerate bilinear form.
The non-Abelian theory is constructed and the resulting sigma 
model is shown to be conformally invariant at the one loop
level.

\section{Duality with non-semi-simple isometries}

The model we would like to consider is 
a Wess-Zumino-Witten (WZW) model defined 
on the group manifold ${\mathcal{M}}_{\mathcal{G}}$. It is based
on the four-dimensional non-semi-simple Lie algebra ${\cal G}$
\cite{nappi} 
\be
\left[J\,,P_i\right]=\epsilon_{ij}P_j\,\,\,,\,\,\,
\left[P_i\,,P_j\right]=\epsilon_{ij}T\,\,\,,\,\,\,
\left[T\,,J\right]=0\,\,\,,\,\,\,
\left[T\,,P_i\right]=0\,\,\,.
\ee
The algebra ${\cal G}$, generated by $T_a=\left\{
P_1,P_2,J,T\right\}$,  has an invariant
bilinear form $\Omega_{ab}$ satisfying (\ref{form}).
Furthermore, it is invertible (there is an inverse matrix 
$\Omega^{ab}$ obeying
$\Omega^{ab}\Omega_{bc}=\delta^a_c$). We have 
\be
\Omega_{ab}=\left(
\begin{array}{cccc}
1&0&0&0\\
0&1&0&0\\
0&0&b&1\\
0&0&1&0
\end{array}\right)\,\,\,\,,\,\,\,\,
\Omega^{ab}=\left(
\begin{array}{cccc}
1&0&0&0\\
0&1&0&0\\
0&0&0&1\\
0&0&1&-b
\end{array}\right)\,\,\,,
\ee
where $b$ is a constant \footnote{String backgrounds based on 
non-semi-simple groups have also been considered in \cite{semi}.}.
\par
In general, given some structure constants $f^a_{ab}$ and their
corresponding non-degenerate and invariant bilinear form 
$\Omega_{ab}$, one constructs the WZW action 
\bea
S\left(g\right)&=&{k\over 8\pi}\int_{\dd \Sigma} {\mathrm{d}}^2x
\sqrt{-\gamma}\gamma^{\mu\nu}\Omega_{ab}B^a_\mu B^b_\nu
+{k\over 24\pi}\int_\Sigma {\mathrm{d}}^3y
\epsilon^{\mu\nu\rho} \Omega_{ad}f^d_{bc} 
B^a_\mu B^b_\nu B^c_\rho \,\,\,\,
\eea
where $g$ is defined on the group manifold ${\mathcal{M}}_{\mathcal{G}}$. 
The quantities $B^a_\mu$ are defined through $B^a_\mu T_a=g^{-1}\dd_\mu g$.
\par
The resulting sigma model is found by choosing an explicit parametrisation
of the group manifold. In the case at hand this is taken to be \cite{nappi}
\be
g=\exp\left(a_1P_1+a_2P_2\right)\exp\left(uJ+vT\right)\,\,\,.
\ee
The gauge-like quantities $B^a_\mu$ are then given by
\bea
B_\mu^1&=&\cos\left(u\right)\dd_\mu a_1 -\sin\left(u\right)
\dd_\mu a_2\nonumber\\
B_\mu^2&=&\cos\left(u\right)\dd_\mu a_2 +\sin\left(u\right)
\dd_\mu a_1\nonumber\\
B_\mu^3&=&\dd_\mu u\nonumber\\ 
B_\mu^4&=&\dd_\mu v +{1\over 2}\epsilon_{ij}
a_j\dd_\mu a_i\,\,\,\,.
\eea
The sigma model reads then
\bea
S\left(g\right)&=&{k\over 8\pi}\int_{\dd \Sigma} {\mathrm{d}}^2x
\sqrt{-\gamma}\gamma^{\mu\nu}\left(\dd_\mu a_i\dd_\nu a_i
+\epsilon_{ij}a_j\dd_\mu a_i\dd_\nu u +
b\dd_\mu u\dd_\nu u  +2\dd_\mu u\dd_\nu v\right)
\nonumber\\
&+&
{k\over 8\pi}\int_{\dd \Sigma} {\mathrm{d}}^2x
\epsilon^{\mu\nu}\left(2u\dd_\mu a_1\dd_\nu a_2\right).
\label{witty}
\eea
One can read off the metric and the antisymmetric tensor field. The dilaton
field vanishes in this case.
\par
The space-time backgrounds  corresponding to (\ref{witty})
satisfy, as expected, the one-loop
conformal invariance conditions
\newpage 
\bea
&&R_{MN} - {1\over 4}H_{MPR}H^{PR}_N +2\nabla_N\nabla_M\Phi=0\nonumber\\
&&\nabla^L\left(e^{-2\Phi}H_{LMN}\right)=0\nonumber\\
&&R -  {1\over 12} H_{MNP}H^{MNP} +4\nabla^N\nabla_N\Phi
-4\dd^N\Phi\dd_N\Phi -{2\over 3\alpha'}\left(c-d\right)=0\,\,\,.
\label{beta}
\eea
Here $M,N,\dots=1,\dots,4$ and correspond to the space-time coordinates
$\varphi^N=\left\{a_1,a_2,u,v\right\}$. 
The central charge $c$ equals to $4$ and is the same
as the dimension of the target space-time $d$. 
\par
Other solutions to the above equations can be generated through
duality transformations.
The WZW action is invariant under the chiral symmetry 
\be
g \longrightarrow LgL^{-1} \,\,\,.
\label{trans}
\ee
Under an infinitesimal chiral transformation 
$L=\alpha^a T_a $ we have
\be
\delta B^a_\mu=\dd_\mu\lambda^a +f^a_{bc}B^b_\mu\lambda^c\,\,\,\,,\,\,\,\,
\lambda^a=-\alpha^a +W^a_b\alpha^b\,\,\,.
\ee
Here the quantity $W^a_b$ is defined via
$W^b_aT_b=g^{-1}T_ag$ and it has the following useful
properties
\bea
\Omega_{ab}W^a_cW^b_d&=&\Omega_{cd}\nonumber\\
\dd_\mu W^a_b&=&f^a_{ec}W^e_bB^c_\mu\nonumber\\
\delta W^a_b&=&\alpha^e\left(f^d_{be}W^a_d -f^a_{ce}W^c_b\right)
\eea
This isometry group is an anomaly-free subgroup and its corresponding
gauged action is written as
\bea
S_1\left(g,A,X\right)&=& S\left(g\right) 
+{k\over 4\pi}\int_{\dd \Sigma}{\mathrm{d}}^2x
\Omega_{ab}\left[P^{\mu\nu}_+B^a_\mu W^b_c A^c_\nu
 -P^{\mu\nu}_-B^a_\mu A^b_\nu
-P^{\mu\nu}_-W^a_cA^c_\mu A^b_\nu \right]
\nonumber\\
&+&{k\over 4\pi}\int_{\dd\Sigma}\epsilon^{\mu\nu}
\Omega_{ab}X^aF^b_{\mu\nu}\,\,\,\,\,.
\eea
The last term is the usual Lagrange multiplier term. 
Since $\Omega_{ab}$ is invertible, the integration over
$X^a$ leads to $F^a_{\mu\nu}=0$ which, owing to gauge invariance,
yields $A^a_\mu=0$. Substituting this back in the gauged action
gives the original WZW theory.
The gauge field
$A_\mu^a$ transforms as in equation (\ref{gaugevar}). We have,
for convenience, defined
the projection matrix
\be
P_{\pm}^{\mu\nu}=\sqrt{-\gamma}\gamma^{\mu\nu}
\pm \epsilon^{\mu\nu}\,\,\,.
\ee
\par
\newpage
The individual infinitesimal coordinate
transformations are written as
\bea
\delta a_1&=&\alpha_{1}\left(1-\cos\left(u\right)\right)
+\alpha_2\sin\left(u\right) -\alpha_3 a_2\nonumber\\
\delta a_2&=&-\alpha_1\sin\left(u\right)
+ \alpha_{2}\left(1-\cos\left(u\right)
\right)
+\alpha_3 a_1\nonumber\\
\delta u &=&0\nonumber\\
\delta v&=&\alpha_{1}\left(-{1\over 2}a_2+{1\over 2}a_1\sin\left(u\right)
+{3\over 2}a_2\cos\left(u\right)\right)\nonumber\\ 
&+&\alpha_{2}\left(-{3\over 2}a_1+{1\over 2}a_1\cos\left(u\right)
-{3\over 2}a_2\sin\left(u\right)\right)
+\alpha_{3}\left(a_2^2-a_1^2\right)
\nonumber\\
\delta X_1&=&\alpha_2 X_3 -\alpha_3 X_2\nonumber\\
\delta X_2&=&-\alpha_1 X_3 +\alpha_3 X_1\nonumber\\
\delta X_3&=&0\nonumber\\
\delta X_4&=&\alpha_1 X_2 -\alpha_2 X_1
\,\,\,.
\eea
The fourth gauge parameter, $\alpha_4$, appears only in the
transformations of the gauge fields $A^a_\mu$.
\par
We are now at a stage where we can eliminate the gauge fields. 
There is, however, a new feature regarding the integration over
the gauge fields. This is due to the presence of the central element
$T$ in the Lie algebra ${\cal G}$. 
In order to see this, the gauged action is written as
\bea
S\left(g,A,X\right)&=& S\left(g\right) 
+{k\over 4\pi}\int_{\dd \Sigma}{\mathrm{d}}^2x\left\{
{1\over 2}\left(\sqrt\gamma\gamma^{\mu\nu}M_{ij}+
\epsilon^{\mu\nu}N_{ij}\right)A^i_\mu A^j_\nu\right.\nonumber\\
&+&
\left[P^{\mu\nu}_+\Omega_{cb}W^b_i B^c_\nu
 -P^{\mu\nu}_-\Omega_{bi}B^b_\nu 
+2\epsilon^{\mu\nu}\Omega_{ib}\dd_\nu X^b\right]A^i_\mu
\nonumber\\
&+&\left.
2\epsilon^{\mu\nu}\left(\dd_\nu X_3 -B^3_\nu\right)A^4_\nu
\right\}\,\,\,\,\,,
\eea
where the indices $i,j=1,2,3$ and we have defined
\bea
M_{ij}&=& 2\Omega_{ij} -\Omega_{ci}W^c_j -\Omega_{cj}W^c_i\nonumber\\
N_{ij}&=& 2\Omega_{cd}X^cf^c_{ij} -\Omega_{ci}W^c_j +\Omega_{cj}W^c_i
\,\,\,\,.
\eea
Notice that the integration over $A^4_\mu$ leads, in the 
path integral, to the constraint
\be 
\epsilon^{\mu\nu}\left(\dd_\nu X_3 -B^3_\nu\right)=0
\label{constraint}
\ee
which can be solved by setting $X_3=u$. This is consistent with
the fact that $\delta u=\delta X_3=0$.
\par
The integration over the rest of the gauge fields is Gaussian and
can be carried out. However, one needs first to choose a gauge fixing
condition.  A suitable gauge is provided by the choice
\be
a_1=\,a_2\,=\,X_1\,=\,0\,\,\,\,.
\ee
When taking into account the constraint (\ref{constraint}),  
the number of fields is equal to four $\varphi^N=
\left\{u,v,X_2,X_4\right\}$.
The matrix in front of the quadratic term in the gauge fields is 
invertible.  Therefore the integration is straitforward and we
get the following non-linear sigma model
\bea 
S&=&{k\over 8\pi}\int_{\dd \Sigma} {\mathrm{d}}^2x
\sqrt{-\gamma}\gamma^{\mu\nu}\left[
b\dd_\mu u\dd_\nu u +2\dd_\mu u \dd_\nu v
-{2\over \cos(u)-1}\dd_\mu X_2\dd_\nu X_2\right.\nonumber\\ 
&+&4{\sin(u)-u\over X_2\left(\cos(u)-1\right)}
\left(\dd_\mu X_2\dd_\nu X_4 - 
\dd_\mu X_2\dd_\nu v\right)\nonumber\\
&+&\left.
2{2u\sin(u)+2\cos(u)-u^2 -2\over X_2^2\left(\cos(u)-1\right)}
\left(\dd_\mu v\dd_\nu v + 
\dd_\mu X_4\dd_\nu X_4 -2\dd_\mu X_4\dd_\nu v\right)
\right]
\,\,\,\,.
\eea
The torsion vanishes in the dual theory. The integration over the
gauge fields leads also to an extra local determinant. The regularisation
of the latter yields a contribution to the dilaton field given by
\cite{buscher,schwarz}
\bea
\Phi&=& -{1\over 2}\ln\left[\det\left(M_{ij}+N_{ij}\right)\right]
       \,\,\, +\,\,\,{\rm{constant}}
\nonumber\\
   &=& -{1\over 2}\ln\left[X_2^2\left(1-\cos(u)\right)\right]
\,\,\, +\,\,\,{\rm{constant}}\,\,\,.
\eea
We have explicitly checked that the resulting space-time backgrounds do 
indeed satisfy the conformal
invariance condition given in (\ref{beta}). The gauge fixing conditons
lead to a non-trivial Faddeev-Popov factor in the path integral measure.
As can be verified, this factor combines with the left-right invariant 
Haar measure, 
${\rm d}a_1{\rm d}a_2{\rm d}u{\rm d}v$, 
of the original WZW model to give the expected measure
of the dual theory 
$e^{-2\Phi}\sqrt{-G}\, {\rm d}X_2{\rm d}X_4{\rm d}u{\rm d}v$ 
\cite{bars}. 
\par
It is worth mentioning that the metric of the original theory describes a
plane wave. This plane is monochromatique and has, in a special coordinate
system, two singularities when $\cos\left(u\right)=\pm 1$. The metric 
of the dual theory
has, in addition to these singularities, a further singularity
at $X_2=0$. The resulting geometry is not that of a palne wave.
A similar model was obtained by considering, in a different context
and using a different group manifold parametrisation, the non-Abelian
dual of the above WZW model \cite{sfetsos}. 

\section{Conclusions}

We have resolved in this paper a standing problem concerning the
implementation of non-Abelian duality based on non-semi-simple
isometry groups. It is shown that the construction of non-Abelian 
dual sigma models is possible only when the isometry group possesses
a non-degenerate invariant bilinear form.  We confirm our analyses by
constructing a non-Abelian dual of a sigma model having non-semi-simple
isometries. The backgrounds of the resulting theory fulfill
the one loop conformal invariance conditons. 
\par
Our analyses can be generalised to other WZW models based on more 
complicated non-semi-groups. Notice also that in the example we
have studied, the metric of the original theory has the translation
symmetry $v\longrightarrow v+\epsilon$.  This translation is generated by 
a null vector. In general, bosonic and supersymmetric string solutions
with covariantly constant null Killing vectors have played a r\^ole
in the construction of dyonic Bogomol'nyi-Prasad-Sommerfield (BPS)
satates \cite{myriam}. 
It would be of interest to examin the effects of non-Abelian
duality on these BPS states. 
\par
Abelian duality has proved to be crucial in the understanding of
string theory and membranes. However, its non-Abelian counterpart
has not been fully explored. One of the areas where non-Abelian 
duality might be of interest is in cosmological and inflationary
models based on string effective theories \cite{cosmo}. 
This is due to the
fact that most of the relevant cosmological models have a tendency
to possess non-Abelian rather than Abelian isometries. 
This is certainly the case in the two examples considered here and which
do not have non-degenerate bilinear forms (Bianchi IV and V). In order
to explore the cosmological implications of non-Abelian duality, one is forced 
to centrally extend the Lie algebras of the two isometries. This would 
provide us with an invertible bilinear form.  Physically, this is  achieved 
by extending the dimension of space-time. This issue is currently under 
investigation.


\end{document}